\documentclass{article}
\usepackage{spconf, amsmath, graphicx, amsfonts, amssymb, amsthm, float}

\title{Phase-Only Analog Encoding for a Multi-Antenna Fusion Center}
\name{Feng Jiang, Jie Chen, and A. Lee Swindlehurst}
\address{Department of Electrical Engineering and Computer Science\thanks{This work is supported by the Air Force Office of Scientific
Research grant FA9550-10-1-0310, and by the National Science Foundation
under grant CCF-0916073.}\\
  University of California at Irvine,\\ Irvine, CA 92697, USA\\
  Email:\{feng.jiang, jie.chen, swindle\}@uci.edu}
  
\begin{document}
%
\maketitle
\begin{abstract}

We consider a distributed sensor network in which the single antenna
sensor nodes observe a deterministic unknown parameter and after
encoding the observed signal with a phase parameter, the sensor nodes
transmit it simultaneously to a multi-antenna fusion center (FC). The
FC optimizes the phase encoding parameter and feeds it back to the
sensor nodes such that the variance of estimation error can be
minimized.  We relax the phase optimization problem to a semidefinite
programming problem and the numerical results show that the
performance of the proposed method is close to the theoretical
bound. Also, asymptotic results show that when the number of sensors
is very large and the variance of the distance between the sensor
nodes and FC is small, multiple antennas do not provide a benefit
compared with a single antenna system; when the number of antennas $M$
is large and the measurement noise at the sensor nodes is small
compared with the additive noise at the FC, the estimation error
variance can be reduced by a factor of $M$.

\end{abstract}
\begin{keywords}
Distributed sensor network, multi-antenna fusion center, maximum likelihood estimation, phase-only analog encoding, asymptotic estimation error.
\end{keywords}
\setlength{\baselineskip}{0.97\baselineskip}
\section{Introduction}
Sensor networks have been widely studied for detection and estimation
problems. Recently, considerable research has focused on the fusion of
analog data in a distributed sensor network to improve estimation
performance. In \cite{Cui:2007}, the authors considered estimation
problems involving single antenna sensors and a single antenna FC. The sensor nodes use an amplify-and-forward scheme to
transmit observations to the FC over fading channels and the optimal
allocation of power to the sensors was investigated. In
\cite{Banavar:2010}, the asymptotic variance of the best linear
unbiased estimator for a single antenna distributed sensor network was
derived and the effect of phase quantization errors were
analyzed. Four different multiple access schemes for a single antenna
decentralized estimation system with a Gaussian source were
investigated in \cite{Leong:2010}, and the scaling laws for large
number of sensors were derived. A coherent multiple access channel was considered in \cite{Xiao:2008} and the optimal linear decentralized estimation scheme was investigated. It is well known that for a multiple
antenna communications system, the link capacity generally increases
linearly with the minimum number of antennas at the transmitter or
receiver.  It is also expected that the estimation performance of a
sensor network would also benefit from a multi-antenna FC, although
prior work on this scenario is limited.  A system with a multi-antenna
FC was considered in \cite{Smith:2009}, which showed that for
Rayleigh fading channels, the reduction in estimation error variance
is bounded by $2$ when the number of sensors approaches infinity.

In this paper we consider a distributed sensor network in which
several single antenna sensor nodes observe a deterministic parameter
corrupted by noise, and simultaneously transmit the observed signal to
a multi-antenna FC. The low-complexity sensor nodes are assumed to
transmit an analog signal with constant power and adjustable phase.
The FC determines the optimal value of the phase for each sensor in
order to minimize the maximum likelihood (ML) estimation error, and
the FC feeds this information back to the sensors so that they can
encode their observed signal accordingly.  The estimation performance
of the phase-optimized sensor network is shown to be considerably
improved compared with sensors that use non-optimized phase.

The paper is organized as follows. Section \ref{sec:two} describes the
system model and Section \ref{sec:three} formulates the phase
optimization problem and proposes a numerical solution based on
semidefinite programming (SDP). Asymptotic analyses for large number
of sensors and large number of antennas are provided in Section
\ref{sec:four}. Numerical results are then presented in Section
\ref{sec:five} and our conclusions are summarized in Section
\ref{sec:six}.

\section{System Model}\label{sec:two}

We assume that $N$ single-antenna sensors in a distributed sensor
network independently observe a deterministic parameter
$\theta\in\mathbb{C}$. The sensor nodes encode the observed signal
with a phase parameter $a_i~(1\le i\le N)$ and simultaneously transmit
it to the FC.  Assuming the FC is configured with $M$ antennas, the
received signal at the FC is expressed as
\begin{equation}
\mathbf{y}=\mathbf{H}\mathbf{a}\theta+\mathbf{H}\mathbf{D}\mathbf{v}+\mathbf{n}\nonumber,
\end{equation}
where $\mathbf{H}=[\mathbf{h}_{1},\dots,\mathbf{h}_{N}]$ and
$\mathbf{h}_{i}\in\mathbb{C}^{M\times 1}$ is the channel vector
between the $i$th sensor and the FC,
$\mathbf{a}=\{a_{1},\dots,a_{N}\}^{T}$ contains the adjustable
phase parameters ($|a_i|=1$), $\mathbf{D}=\mathrm{diag}\{a_{1},\dots,a_{N}\}$,
$\mathbf{v}$ is the measurement noise at the sensor nodes which is
assumed to be Gaussian distributed with covariance
$\mathbf{V}=\mathbb{E}\{\mathbf{v}\mathbf{v}^{H}\}=\mathrm{diag}\left\{\sigma_{v,1}^2,\cdots,\sigma_{v,N}^2\right\}$,
$\mathbf{n}$ is the additive Gaussian noise at the FC with
covariance
$\mathbb{E}\{\mathbf{n}\mathbf{n}^{H}\}=\sigma_{n}^2\mathbf{I}_{M}$
and $\mathbf{I}_{M}$ is an $M\times M$ identity matrix.  

Assuming the FC is aware of the channel matrix 
$\mathbf{H}$, the noise covariance $\mathbf{V}$ and $\sigma_n^2$,
the FC calculates the ML estimate of $\theta$ using \cite{Kay:1993}
\begin{equation}
\hat{\theta}_{ML}=\frac{\mathbf{a}^{H}\mathbf{H}^{H}(\mathbf{H}\mathbf{V}\mathbf{H}^{H}+\sigma_{n}^2\mathbf{I}_{M})^{-1}\mathbf{y}}{\mathbf{a}^{H}\mathbf{H}^{H}(\mathbf{H}\mathbf{V}\mathbf{H}^{H}+\sigma_{n}^2\mathbf{I}_{M})^{-1}\mathbf{H}\mathbf{a}}.\nonumber
\end{equation}
The estimator $\hat{\theta}_{ML}$ is unbiased and the variance of
$\hat{\theta}_{ML}$ is given by
\begin{equation}\label{eq:ml}
\mathrm{Var}(\hat{\theta}_{ML})=\left(\mathbf{a}^{H}\mathbf{H}^{H}(\mathbf{H}\mathbf{V}\mathbf{H}^{H}+\sigma_{n}^2\mathbf{I}_{M})^{-1}\mathbf{H}\mathbf{a}\right)^{-1}.
\end{equation}
The variance is lower bounded by
\begin{equation}\label{eq:lb}
\mathrm{Var}(\hat{\theta}_{ML})\!\ge\!\frac{1}{N\lambda_{\max}\left(\mathbf{H}^{H}(\mathbf{H}\mathbf{V}\mathbf{H}^{H}+\sigma_{n}^2\mathbf{I}_{M})^{-1}\mathbf{H}\right)},
\end{equation}
where $\lambda_{\max}(\cdot)$ denotes the largest eigenvalue of a matrix.

\section{Phase-only Analog Encoding Method}\label{sec:three}
From (\ref{eq:ml}), we see that
$\mathrm{Var}(\hat{\theta}|\mathbf{H})_{ML}$ is inversely proportional
to a quadratic form in $\mathbf{a}$, which suggests the possibility of 
minimizing the variance by adjusting the phase of the signals transmitted
by the sensors. The advantage of adjusting only the phase is that such
adjustments do not impact the covariance of the sensor noise observed
at the FC, unlike what would occur if the transmit power of the sensors
was also adjusted.  This is the reason for the assumption of sensors
with constant transmit power.

We formulate the following optimization problem at the FC:
\begin{eqnarray}\label{eq:opt}
\min_{\mathbf{a}} &&\mathrm{Var}(\hat{\theta}_{ML})\\
s. t. &&|a_i|=1,\; i=1, \dots, N.\nonumber
\end{eqnarray}
Define
$\mathbf{B}=\mathbf{H}^{H}(\mathbf{H}\mathbf{V}\mathbf{H}^{H}+\sigma_{n}^2\mathbf{I}_{M})^{-1}\mathbf{H}$,
then we can rewrite problem (\ref{eq:opt}) as
\begin{eqnarray}\label{eq:quardra}
\max_{\mathbf{a}} &&\mathbf{a}^{H}\mathbf{B}\mathbf{a}\\
s. t. &&|a_i|=1,\; i=1, \dots, N.\nonumber
\end{eqnarray}
If there are only two sensors in the network, a closed-form solution to 
problem~(\ref{eq:quardra}) can be obtained. 
Defining $\mathbf{B}=\left[\begin{array}{cc} c&be^{j\beta}\\
be^{-j\beta}&c\end{array}\right]$ with $b,c>0$, then the largest eigenvalue
of $\mathbf{B}$ is given by
\begin{equation}
\lambda_{\max}(\mathbf{B})=c+b,\nonumber
\end{equation}
and the solution to~(\ref{eq:quardra}) is given by the corresponding eigenvector 
\begin{equation}
\mathbf{a}=\left[e^{j\beta_1}, e^{j\beta_2}\right]^T,\nonumber
\end{equation}
where $\beta_1-\beta_2=\beta$.

For the general situation where $N>2$, we relax
problem~(\ref{eq:quardra}) and convert it to a standard SDP problem
that can be solved with standard tools such as {\tt cvx}
\cite{cvx:2011}. To begin with, we rewrite~(\ref{eq:quardra}) into the
following equivalent form
\begin{eqnarray}\label{eq:rankone}
\max_{\mathbf{A}} && \mathrm{tr}(\mathbf{B}\mathbf{A})\\
s. t. &&\mathbf{A}_{i,i}=1,\; i=1, \dots, N\nonumber\\
      && \mathrm{rank}(\mathbf{A})=1\nonumber\\
      &&\mathbf{A}\succeq 0\nonumber,
\end{eqnarray}
where $\mathbf{A}_{i,i}$ denotes $(i,i)$th element of
$\mathbf{A}$. Relaxing the rank-one constraint on $\mathbf{A}$, we can
convert problem (\ref{eq:rankone}) to a standard SDP problem:
\begin{eqnarray}\label{eq:sdp}
\max_{\mathbf{A}} && \mathrm{tr}(\mathbf{B}\mathbf{A})\\
s. t. &&\mathbf{A}_{i,i}=1,\; i=1, \dots, N\nonumber\\
      &&\mathbf{A}\succeq 0\nonumber.
\end{eqnarray}
Defining $\mathbf{B}_r=\mbox{\rm real}\{\mathbf{B}\}$,
$\mathbf{B}_i=\mbox{\rm imag}\{\mathbf{B}\}$, and similarly for
$\mathbf{A}_r$ and $\mathbf{A}_i$, (\ref{eq:sdp}) can be converted to
the equivalent real form
\begin{eqnarray}\label{eq:realsdp}
\max_{\{\mathbf{A}_r,\mathbf{A}_i\}} && \mathrm{tr}(\mathbf{B}_r\mathbf{A}_r-\mathbf{B}_i\mathbf{A}_i)\\
s. t. &&\mathbf{A}_{r~i,i}=1,\; i=1, \dots, N\nonumber\\
      &&\left[\begin{array}{cc} \mathbf{A}_r&-\mathbf{A}_i\\
                \mathbf{A}_i&\mathbf{A}_r\end{array}\right]\succeq 0\nonumber.
\end{eqnarray}
Problem (\ref{eq:realsdp}) can be efficiently solved using a
standard interior point method.  Denote the solution to problem (\ref{eq:realsdp}) as
$\mathbf{A}^{*}$. If $\mathrm{rank}(\mathbf{A}^{*})>1$, we can use a
method similar to Algorithm 2 in \cite{Zhang:2011} to extract a
rank-one solution. In Section \ref{sec:five}, the numerical results show that
the performance of the proposed phase encoding method is close to the
theoretical lower bound in~(\ref{eq:lb}).

\section{Asymptotic Analysis}\label{sec:four}
In the asymptotic analyses conducted in this section, we assume a
path-loss based model
\begin{equation}
\mathbf{h}_{i}=\frac{1}{d_i^\alpha}\tilde{\mathbf{h}}_{i},\nonumber
\end{equation}
where $d_i$ denotes the distance between the $i$th sensor and the FC,
$\alpha$ is the path loss exponent and $\tilde{\mathbf{h}}_i$ denotes
the normalized channel component.  Furthermore, the elements of the
normalized channel are assumed to have constant amplitude and random
phase:
\begin{equation}
\tilde{\mathbf{h}}_{i}=[e^{j\gamma_{i,1}}, e^{j\gamma_{i,2}},\dots, e^{j\gamma_{i,M}}]^{T}\nonumber,
\end{equation}
where $\gamma_{i,j}$ is assumed to be uniformly distributed over $\left[0,2\pi\right]$.

\subsection{Performance Bound for Large Number of Sensors}
From (\ref{eq:lb}), the lower bound of $\mathrm{Var}(\hat{\theta}|\mathbf{H})_{ML}$ depends on the largest eigenvalue of $\mathbf{H}^{H}(\mathbf{H}\mathbf{V}\mathbf{H}^{H}+\sigma_{n}^2\mathbf{I}_{M})^{-1}\mathbf{H}$. To evaluate this eigenvalue, we approximate $\mathbf{H}\mathbf{V}\mathbf{H}^H$ as a diagonal matrix. The $(m,n)$th element of $\mathbf{H}\mathbf{V}\mathbf{H}^H$ is given by
\begin{eqnarray}
\mathbf{H}\mathbf{V}\mathbf{H}^{H}_{m,n}=\sum_{i=1}^{N}\frac{e^{j(\gamma_{i,m}-\gamma_{i,n})}\sigma_{v,i}^2}{d_i^{2\alpha}}.\nonumber
\end{eqnarray}
According to the strong law of large numbers, when $N\to \infty$, the following equation holds
\begin{eqnarray}\label{eq:mean}
\lim_{N\to\infty}\frac{1}{N}\!\!\sum_{i=1}^{N}\!\frac{e^{j(\gamma_{i,m}\!-\gamma_{i,n})}\sigma_{v,i}^2}{d_i^{2\alpha}}&\!\!\!\!\overset{(a)}{=}\!\!\!\!\!&\!\mathbb{E}\!\left\{\!\!\frac{\sigma_{v,i}^2}{d_i^{2\alpha}\!}\!\right\}\!\mathbb{E}\left\{\!e^{j(\gamma_{i,m}\!-\gamma_{i,n})}\!\right\}\nonumber, \\
&\!\!\!\overset{(b)}{=}\!\!\!\!&\left\{\begin{array}{ll}
\mathbb{E}\!\left\{\!\frac{\sigma_{v,i}^2}{d_i^{2\alpha}\!}\right\}, &m=n,\\
0,&m\ne n,
\end{array}\right.
\end{eqnarray}
where (a) follows from the assumption that $\gamma_{i,m}$, $d_i$ and
$\sigma_{v,i}^2$ are independent of each other and (b) is due to the fact that
$\gamma_{i,m}$ and $\gamma_{i,n}$ are independent and uniformly
distributed over $[0,2\pi]$. Thus, when $N\to\infty$, we can
approximate $\mathbf{H}\mathbf{V}\mathbf{H}^{H}$ as
\begin{equation}\label{eq:approx}
\lim_{N\to\infty}\mathbf{H}\mathbf{V}\mathbf{H}^{H}\approx N\mathbb{E}\left\{\frac{\sigma_{v,i}^2}{d_i^{2\alpha}}\right\}\mathbf{I}_{M},
\end{equation}
where $\mathbb{E}\left\{\frac{\sigma_{v,i}^2}{d_i^{2\alpha}}\right\}=\frac{1}{N}\sum_{i=1}^{N}\frac{\sigma_{v,i}^2}{d_i^{2\alpha}}$. Based
on~(\ref{eq:approx}), we have
\begin{eqnarray}\label{eq:lambdamax}
&&\lambda_{\max}\left(\mathbf{H}^{H}(\mathbf{H}\mathbf{V}\mathbf{H}^{H}+\sigma_{n}^2\mathbf{I}_{M})^{-1}\mathbf{H}\right)\nonumber\\
&\approx&\frac{1}{N\mathbb{E}\left\{\frac{\sigma_{v,i}^2}{d_i^{2\alpha}}\right\}+\sigma_n^2}\lambda_{\max}(\mathbf{H}^H\mathbf{H})\nonumber\\
&\overset{(c)}{\approx}&\frac{N\mathbb{E}\left\{\frac{1}{d_i^{2\alpha}}\right\}}{N\mathbb{E}\left\{\frac{\sigma_{v,i}^2}{d_i^{2\alpha}}\right\}+\sigma_n^2},
\end{eqnarray}
where (c) is due to
$\lambda_{\max}(\mathbf{H}^{H}\mathbf{H})=\lambda_{\max}(\mathbf{H}\mathbf{H}^{H})$. Plugging
(\ref{eq:lambdamax}) into (\ref{eq:lb}), the following lower bound for
$\mathrm{Var}(\hat{\theta}_{ML})$ can be obtained:
\begin{equation}\label{eq:lb2}
\lim_{N\to\infty}\mathrm{Var}(\hat{\theta}_{ML})\ge\frac{\sigma_{n}^2\!\!+\!\sum_{i=1}^{N}\frac{\sigma_{v,i}^2}{d_i^{2\alpha}}}{N\sum_{i=1}^{N}\frac{1}{d_i^{2\alpha}}}.
\end{equation}

Clearly, an upper bound on the estimate variance can be found by
considering the single-antenna case, which results in 
\begin{equation}\label{eq:allone}
\lim_{N\to\infty}\mathrm{Var}(\hat{\theta}_{ML})\le\frac{\sigma_{n}^2+\sum_{i=1}^{N}\frac{\sigma_{v,i}^2}{d_i^{2\alpha}}}{\left(\sum_{i=1}^{N}\frac{1}{d_i^\alpha}\right)^2}.
\end{equation}
When $N\to\infty$, the ratio of the lower to the upper bound is 
given by
\begin{eqnarray}\label{eq:ratio}
\lim_{N\to\infty}\frac{\left(\sum_{i=1}^{N}\frac{1}{d_i^\alpha}\right)^2}{N\sum_{i=1}^{N}\frac{1}{d_i^{2\alpha}}}=\frac{\left(\mathbb{E}\left\{\frac{1}{d_{i}^\alpha}\right\}\right)^2}{\mathbb{E}\left\{\frac{1}{d_{i}^{2\alpha}}\right\}}=1-\frac{\mathrm{Var}\left\{\frac{1}{d_{i}^\alpha}\right\}}{\mathbb{E}\left\{\frac{1}{d_{i}^{2\alpha}}\right\}}.
\end{eqnarray}
Interestingly, we see from (\ref{eq:ratio}) that when
$\mathrm{Var}\left\{\frac{1}{d_{i}^\alpha}\right\}\ll
\mathbb{E}\left\{\frac{1}{d_{i}^{2\alpha}}\right\}$, the gap between
the bounds is small and the availability of multiple antennas at the
FC does not provide a benefit compared with the single antenna system.

\subsection{Scaling Law with Number of Antennas}
Using the matrix inversion lemma, the inverse of noise covariance matrix
can be written as
\begin{eqnarray}\label{eq:eigen}
&&\vspace{-5em}\!\!\!\!\mathbf{H}^{H}(\mathbf{H}\mathbf{V}\mathbf{H}^{H}+\sigma_{n}^2\mathbf{I}_{M})^{-1}\mathbf{H}\nonumber\\
&{\!\!=\!\!}\!\!&\!\!\!\mathbf{H}^{H}\left(\!\frac{1}{\sigma_n^2}\mathbf{I}_{M}\!-\!\frac{1}{\sigma_n^4}\mathbf{H}\left(\mathbf{V}^{-1}\!\!+\!\!\frac{1}{\sigma_n^2}\mathbf{H}^{H}\mathbf{H}\right)^{-1}\mathbf{H}^{H}\right)\mathbf{H}\nonumber\\
&=\!\!&\!\!\!\frac{1}{\sigma_n^2}\mathbf{H}^{H}\!\mathbf{H}\!-\!\frac{1}{\sigma_n^4}\mathbf{H}^{H}\!\mathbf{H}\left(\!\mathbf{V}^{-1}\!\!+\!\frac{1}{\sigma_n^2}\mathbf{H}^{H}\!\mathbf{H}\!\right)^{\!\!\!-1}\!\!\!\mathbf{H}^{H}\!\mathbf{H}.\end{eqnarray}
The $(m,n)$th element of $\mathbf{H}^{H}\mathbf{H}$ is given by
\begin{equation}
\mathbf{H}^{H}\mathbf{H}_{m,n}=\frac{1}{d_m^\alpha d_n^{\alpha}}\sum_{i=1}^{M}e^{j\left(\gamma_{n,i}-\gamma_{m,i}\right)}.\nonumber
\end{equation}
Similar to (\ref{eq:mean}), when $M\to\infty$, the following holds:
\begin{equation}\label{eq:mean2}
\lim_{M\to\infty}\frac{1}{M}\sum_{i=1}^{M}e^{j\left(\gamma_{n,i}-\gamma_{m,i}\right)}=\left\{\begin{array}{ll}
1, &m=n,\\
0,&m\ne n.
\end{array}\right.
\end{equation}
Based on (\ref{eq:mean2}), we can approximate $\mathbf{H}^{H}\mathbf{H}$ as
\begin{eqnarray}\label{eq:approx2}
\lim_{M\to\infty}\mathbf{H}^{H}\mathbf{H}\approx M\mathrm{diag}\left\{\frac{1}{d_1^{2\alpha}},\cdots, \frac{1}{d_N^{2\alpha}}\right\}.
\end{eqnarray}
Substituting (\ref{eq:approx2}) into (\ref{eq:eigen}), we have   
{\begin{eqnarray}\label{eq:eigen2}
&&\mathbf{H}^{H}(\mathbf{H}\mathbf{V}\mathbf{H}^{H}+\sigma_{n}^2\mathbf{I}_{M})^{-1}\mathbf{H}\nonumber\\
&\approx&\mathrm{diag}\left\{\frac{M}{d_1^{2\alpha}\sigma_n^2+M\sigma_{v,i}^2},\cdots,\frac{M}{d_N^{2\alpha}\sigma_n^2+M\sigma_{v,N}^2}\right\}\nonumber.
\end{eqnarray}}
\!\!Since $\mathbf{H}^{H}(\mathbf{H}\mathbf{V}\mathbf{H}^{H}+\sigma_{n}^2\mathbf{I}_{M})^{-1}\mathbf{H}$ is thus approximately diagonal, for any phase vector $\mathbf{a}$, we have 
\begin{equation}\label{eq:varapprox}
\lim_{M\to\infty}\mathrm{Var}(\hat{\theta}_{ML})\approx\frac{1}{M\sum_{i=1}^{N}\frac{1}{d_i^{2\alpha}\sigma_n^2+M\sigma_{v,i}^2}}.
\end{equation}
From (\ref{eq:varapprox}), it can be observed that when
$M\sigma_{v,i}^2\ll\sigma_n^2d_i^2$, a reduction in the
estimate variance by a factor of $M$ can approximately be
achieved, and to realize this gain, the phase vector
$\mathbf{a}$ can be selected arbitrarily.

\section{Numerical Results}\label{sec:five}
To evaluate the performance of the proposed approach, several
numerical experiments were carried out. In the numerical examples, the
transmit power of sensor nodes are normalized to 1 and the path loss
exponent $\alpha$ is set to $1$. Both $d_i$ and $\sigma_{v,i}^2$ are
assumed to be uniformly distributed, with $d_i$
ranging from $2$ to $7$, $\sigma_{v,i}^2$ between $0.001$ to $0.01$ and the
additive noise power $\sigma_n^2$ is set to $0.1$. In the results,
each point on the curve is obtained by averaging over $300$
realizations of $\tilde{\mathbf{h}}_{i}$. To evaluate the performance
without feedback, $\mathbf{a}$ is set to a vector of all ones (the random
phase component is subsumed in the channel vector).  The results in
Fig.~1 show that the performance of the proposed method is very close
to the lower bound in eq.~(\ref{eq:lb}) and significantly better than
without feedback.  Fig.~2 shows that the benefit of the feedback
reduces as the number of FC antennas grows, although the gain is still
significant for reasonable values of $M$ below 10.  We see in both
figures that as either $N$ or $M$ get large, our asymptotic expressions
match the numerical results well.

\begin{figure}
\centering
\centerline{\includegraphics[height=2.3in, width=3in]{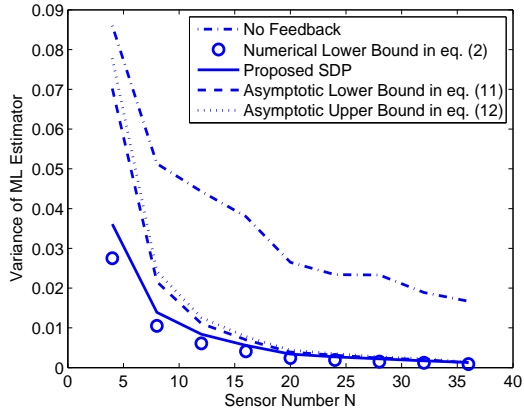}}
\caption{Performance of the ML estimator with increasing number of sensors for $M=4$.}
\end{figure}

\begin{figure}
\centering
\centerline{\includegraphics[height=2.3in, width=3in]{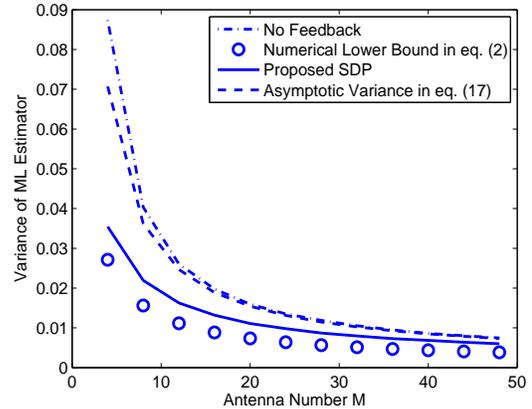}}
\caption{Performance of the ML estimator with increasing number of antennas for $N=4$.}
\end{figure}

\section{Conclusion}\label{sec:six}
In this paper, we investigated a phase-only analog encoding scheme for
a distributed sensor network composed of single-antenna sensors and a
multi-antenna FC. We relaxed the phase optimization problem to an SDP
and the numerical results show that the performance of the proposed
method is close to the theoretical lower bound. Also, asymptotic
results for cases with large numbers of sensors or antennas were
derived. The asymptotic results indicate that when the number of
senors is large and the variance of the distance between the sensor
nodes and the FC is small, the availability of multiple antennas at
the FC does not provide a benefit compared with the single antenna
case.  On the other hand, when the number of antennas $M$ is large and
the measurement noise and the additive noise satisfy
$M\sigma_{v,i}^2\ll\sigma_n^2d_i^{2\alpha}$, a reduction in estimation
variance of $M$ can be obtained.




\bibliographystyle{IEEEbib}
\bibliography{reference}
\end{document}